\documentclass[aps,prl,twocolumn,
preprintnumbers,
amsmath,amssymb,showkeys,floatfix,nofootinbib,reprint]{revtex4-1}

\usepackage{amsmath}
\usepackage{amsfonts}
\usepackage{amssymb}
\usepackage{graphicx}
\usepackage{color}
\usepackage{bm}
\usepackage{hyperref}
\usepackage{diagbox}

\begin{document}

%
%
%

\def\ada#1{\textcolor{blue}{#1}}
\def\jonas#1{\textcolor{red}{#1}}

\def\ket#1{ $ \left\vert  #1   \right\rangle $}
\def\ketm#1{  \left\vert  #1   \right\rangle   }
\def\bra#1{ $ \left\langle  #1   \right\vert $ }
\def\bram#1{  \left\langle  #1   \right\vert   }
\def\spr#1#2{ $ \left\langle #1 \left\vert \right. #2 \right\rangle $ }
\def\sprm#1#2{  \left\langle #1 \left\vert \right. #2 \right\rangle   }
\def\me#1#2#3{ $ \left\langle #1 \left\vert  #2 \right\vert #3 \right\rangle $}
\def\mem#1#2#3{  \left\langle #1 \left\vert  #2 \right\vert #3 \right\rangle   }
\def\redme#1#2#3{ $ \left\langle #1 \left\Vert
                  #2 \right\Vert #3 \right\rangle $ }
\def\redmem#1#2#3{  \left\langle #1 \left\Vert
                  #2 \right\Vert #3 \right\rangle   }
\def\threej#1#2#3#4#5#6{ $ \left( \matrix{ #1 & #2 & #3  \cr
                                           #4 & #5 & #6  } \right) $ }
\def\threejm#1#2#3#4#5#6{  \left( \matrix{ #1 & #2 & #3  \cr
                                           #4 & #5 & #6  } \right)   }
\def\sixj#1#2#3#4#5#6{ $ \left\{ \matrix{ #1 & #2 & #3  \cr
                                          #4 & #5 & #6  } \right\} $ }
\def\sixjm#1#2#3#4#5#6{  \left\{ \matrix{ #1 & #2 & #3  \cr
                                          #4 & #5 & #6  } \right\} }

\def\ninejm#1#2#3#4#5#6#7#8#9{  \left\{ \matrix{ #1 & #2 & #3  \cr
                                                 #4 & #5 & #6  \cr
                         #7 & #8 & #9  } \right\}   }
%
%
%
%

%
%

\title{$^{93m}$Mo isomer depletion via beam-based  nuclear excitation by electron capture}

%
%

\author{Yuanbin \surname{Wu}}
\email{yuanbin.wu@mpi-hd.mpg.de}
\affiliation{Max-Planck-Institut f\"ur Kernphysik, Saupfercheckweg 1, D-69117 Heidelberg, Germany}

\author{Christoph H. \surname{Keitel}}
\affiliation{Max-Planck-Institut f\"ur Kernphysik, Saupfercheckweg 1, D-69117 Heidelberg, Germany}

\author{Adriana \surname{P\'alffy}}
\email{Palffy@mpi-hd.mpg.de}
\affiliation{Max-Planck-Institut f\"ur Kernphysik, Saupfercheckweg 1, D-69117 Heidelberg, Germany}

\date{\today}

%
%
%
%
%
%
%
\begin{abstract}
A recent nuclear physics experiment [C. J. Chiara {\it et al.}, Nature (London) {\bf 554}, 216 (2018)]  reports the first direct observation of nuclear excitation by electron capture (NEEC) in the depletion of the  $^{93m}$Mo isomer. The experiment used a beam-based setup in which Mo highly charged ions with nuclei in the isomeric state $^{93m}$Mo at 2.4 MeV excitation energy were slowed down in a solid-state target. In this process, nuclear excitation to a higher triggering level  led to isomer depletion.  The reported excitation  probability $P_{\rm{exc}} = 0.01$ was solely attributed to the so-far unobserved process of NEEC in lack of a different known channel of comparable efficiency. In this work, we investigate theoretically the beam-based setup and calculate excitation rates via NEEC using state-of-the-art atomic structure  and ion stopping power models. For all scenarios, our results disagree with the experimental data by approximately nine orders of magnitude. This stands in conflict with the conclusion that NEEC was the excitation mechanism behind the observed depletion rate.

\end{abstract}

\maketitle


In nuclear physics, the term {\it isomer} denotes a long-lived excited nuclear state. Isomers  pose challenging  riddles to nuclear structure theory \cite{WalkerN1999,DracoulisRPP2016} and may play a significant role for nucleosynthesis in astrophysical plasmas \cite{Reifarth2018}.  In  terrestrial laboratories one hopes to achieve control of isomeric state population to design novel energy storage solutions. Isomer depletion refers to the core idea behind such energy storage: excitation of the nuclear isomer to a higher lying so-called triggering state together with an advantageous decay branching ratio thereof can lead to the controlled release on demand of the stored nuclear energy \cite{WalkerN1999, AprahamianNP2005, BelicPRL1999, CollinsPRL1999, BelicPRC2002, CarrollLPL2004, PalffyPRL2007, ZadernovskyHI2002}.  Excitation can occur over several channels,  by photoabsorption, Coulomb excitation, inelastic scattering, or coupling to the electronic shell.

\begin{figure}
\centering
\includegraphics[width=1.0\linewidth]{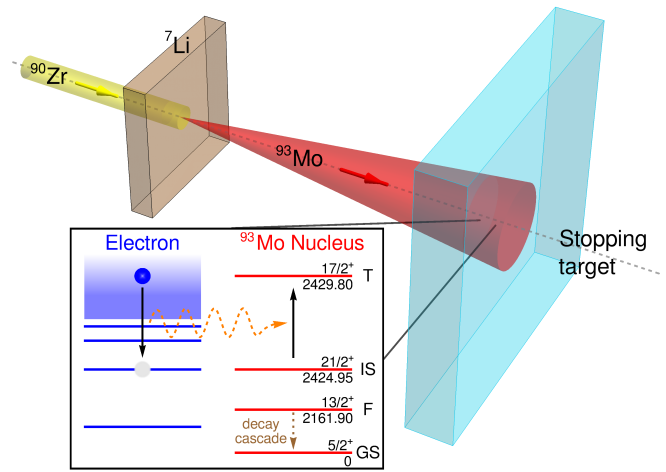}%
\caption{ Sketch of  the beam-based experimental setup with a graphical illustration of the NEEC process and the partial level scheme of  $^{93m}$Mo. In Ref.~\cite{ChiaraNature2018}, the stopping target comprised a C and a Pb layer.
The nuclear isomeric ($IS$), triggering ($T$), intermediate ($F$) and ground state ($GS$) levels are labeled by their spin, parity and energy in keV, taken from Ref.~\cite{ensdf}. Energy intervals are not to scale.
\label{fig:setupb}}
\end{figure}

Nuclear excitation by electron capture (NEEC) is one of these possible excitation mechanisms \cite{PalffyPRL2007}. This process is the time-reversed internal conversion (IC) and occurs when electron recombination into the atomic shell at the exact resonance energy excites the nucleus \cite{GoldanskiiPLB1976, PalffyCP2010}. Theoretically, NEEC has been investigated for channeling through crystals \cite{Cue1989,Kimball1,Kimball2}, in laser-generated plasmas \cite{HarstonPRC1999, GosselinPRC2004, GosselinPRC2007, GobetNIMPRSA2011, GobetNIMPRSA2011, DracoulisRPP2016, GunstPRL2014, GunstPOP2015, WuPRL2018, GunstPRE2018} or in storage ring scenarios \cite{PalffyPRA2006,PalffyPLB2008}. State-of-the-art NEEC theory was so far benchmarked using the data available on its inverse process IC \cite{PalffyPRL2007, PalffyPhD2006, BilousPRA2017}. The first experimental evidence of NEEC was only recently reported in the isomer depletion of the 2.4 MeV $^{93m}$Mo isomer (half-life 6.85 h)  in a beam-based setup  \cite{ChiaraNature2018}.  Fast recoiled $^{93m}$Mo isomers were produced via nuclear reactions in collisions of a 840 MeV $^{90}$Zr beam on a $^7$Li target. This secondary isomeric beam then reached a stopping target comprising a thin carbon layer backed with $^{208}$Pb, as illustrated in Fig.~\ref{fig:setupb}. In the stopping process, Mo ions were stripped of electrons which recombined back later on upon ion deceleration. Provided the resonance condition in the rest frame of the ion was fulfilled,  NEEC   depleted the isomer by driving the 4.85 keV electric quadrupole ($E2$) transition from the isomer to a triggering level $T$ (see partial level scheme in Fig.~\ref{fig:setupb}) which subsequently decayed via a cascade to the ground state.


Ref.~\cite{ChiaraNature2018} reports the clear signal of isomer depletion observed by direct measurement of the  268 keV gamma-ray photon emitted in the transition from state $T$ to state $F$ below the isomer.  The depletion probability $P_{\rm{exc}} = 0.01$  per $^{93m}$Mo  was extracted from the experimental data. Since the branching ratio of the 268 keV transition from $T$ to $F$ equals unity, this probability is both the isomer depletion probability and the nuclear excitation probability from the isomeric state to the triggering level $T$. 
The direct experimental evidence does not point at any nuclear excitation mechanism in particular, and only confirms the depletion of the $^{93m}$Mo isomer. Ref.~\cite{ChiaraNature2018} carefully checked that the signal is not due to contaminant reactions and also provided theoretical estimates on Coulomb excitation and inelastic scattering, which yielded much smaller probabilities than the observed one. However, 
 theoretical NEEC rates for the experimental setting were not provided. With previous works on beam-based setups \cite{KaramianPAN2012, PolasikPRC2017} giving only qualitative arguments, the question on the magnitude of NEEC for $^{93m}$Mo isomer depletion remained unanswered from theory side.

It is the purpose of this Letter to provide the missing theoretical study of a beam-based scenario for NEEC as depleting mechanism for $^{93m}$Mo. Our analysis models the ion charge state distribution of the $^{93m}$Mo isomers, the ion deceleration  and the NEEC process using state-of-the-art atomic structure calculations and several stopping power models. For all considered models, we obtain NEEC probabilities of approx. $10^{-11}$,  nine orders of magnitude smaller than the experimental value $P_{\rm{exc}} = 0.01$ reported in Ref.~\citep{ChiaraNature2018}. A comparison between different stopping power models and consistency checks with  another NEEC scenario available in the literature support the obtained values. The tremendous difference between the theoretical and experimental isomer depletion probabilities speaks against NEEC as the underlying nuclear excitation mechanism for the isomer depletion observed in Ref.~\citep{ChiaraNature2018}. Our findings  support further investigations of the so far considered experimental and theoretical channels, and the search for new possible depletion mechanisms.


The total theoretical probability per $^{93m}$Mo ion in the beam-target scenario is given by the sum of NEEC probabilities over all the possible recombination channels (orbitals) $\alpha$,  and over the  entire charge state distribution of the incoming ions,  integrated over the interaction time
\begin{equation}
  P = \sum_{q, \alpha}\int  f_{q} \phi \sigma_{q}^{\alpha} dt \label{P_neec}\, ,
\end{equation}
where $f_{q}$ is the  ion fraction in charge state $q$, 
$\phi$ the  electron flux in the rest frame of the ion, and $
\sigma_{q}^{\alpha}$ the NEEC cross section into channel $\alpha$ for an initial ion with charge state $q$, respectively. These quantities depend indirectly on time via the varying ion energy in the deceleration process. A change of variable can be made  \cite{WeiPRC2013, AikawaNIMPRSB2015} by introducing the stopping power through the material $-d E^{ \mathrm {ion}} /dx$, which determines the time-dependent ion velocity and correspondingly (in the rest frame of the Mo ions)  the electron recombination energy for the NEEC process. 
The resonant NEEC cross section depends on the recombining continuum electron energy $E$ primarily via a normalized Lorentz profile,
\begin{equation}
  \sigma_{q}^{\alpha} (E) =  S_{q}^{\alpha}(E) \frac{\Gamma_{q, \alpha}/(2\pi)}{(E-E_{q, \alpha})^2 + \frac{1}{4} \Gamma_{q, \alpha}^2}\, ,
\end{equation}
where $S_{q}^{\alpha}(E)$ is the NEEC resonance strength, only slowly varying 
with respect to the electron energy,
and $E_{q, \alpha}$ and  $\Gamma_{q, \alpha}$ are the recombining electron energy and the natural width of the resonant state, respectively. The continuum electron energy is given by the difference between the nuclear transition energy 4.85 keV and the electronic energy transferred to the bound atomic shell in the recombination process. For NEEC into the electronic ground state,  $\Gamma_{q, \alpha}$ is given by the nuclear state width and is  $10^{-7}$ eV for the 2429.80 keV level $T$ above the isomer \cite{ensdf}. The Lorentz profile can then be approximated by a Dirac-delta function. However, if the electron recombination occurs into an excited electronic configuration, the width of the Lorentz profile is determined by the electronic width, typically on the order of 1 eV \cite{AtomicTables}. This  value is still  small compared to the continuum electron energies of  few keV.

In order to relate the electronic and ion energies, we assume that the electron temperature in the solid-state target is very small, so that we can neglect the electron velocity in the laboratory frame. For the C target this is well justified and only introduces deviations of few precent compared to the more accurate treatment based on the Thomas-Fermi approximation \cite{Kimball1,Kimball2}. This leads to the relation $E=(m_e/m_i)E^{\mathrm {ion}}$, where $m_e$ and $m_i$ are the electron and ion masses, respectively.
The very narrow Lorentz profile in the expression of the NEEC cross section justifies considering the ion fraction $f_{q}$, the electron flux $\phi$, the stopping power $- d E^{ \mathrm {ion}} /dx$, and the NEEC resonance strength $S_{q}^{\alpha}$ to be constant for the narrow energy interval of the resonance $\Gamma_{q, \alpha}$. This approximation has a relative accuracy of $\Gamma_{q, \alpha}/E_{q, \alpha}$, i.e., it is for all practical purposes exact for NEEC into ground state configurations for which $\Gamma_{q, \alpha}\approx 10^{-7}$ eV and has a relative accuracy of $10^{-3}$ for the case that the capture occurs into excited electronic states and $\Gamma_{q, \alpha}\lessapprox 1$ eV. Performing the energy integration over the normalized Lorentz profile, we obtain the total NEEC probability 
\begin{eqnarray}
  P &=& \sum_{q, \alpha}  f_{q}(E^{ \mathrm {ion}}_{q, \alpha})\, n_e\, S_{q}^{\alpha}(E_{q, \alpha}) \frac{m_i}{m_e}\frac{1}{-\left.\left(dE^{ \mathrm {ion}}/dx \right)\right|_{E^{ \mathrm {ion}}_{q, \alpha}}}\nonumber \\
  &\equiv& \sum_{q, \alpha}  f_{q}(E^{ \mathrm {ion}}_{q, \alpha}) P_{q}^{\alpha}(E^{ \mathrm {ion}}_{q, \alpha})\, ,
  \label{total_P}
\end{eqnarray}
where $n_e$ is the  electron density, for which we consider the solid-state value to obtain an upper limit estimate.  Furthermore, $P_{q}^{\alpha}$ denotes the NEEC probability  into channel $\alpha$ for an initial ion with charge state $q$.

We calculate the NEEC cross sections following the formalism first developed in Ref.~\cite{PalffyPRA2006} and later used for a number of NEEC studies for highly charged ions or plasmas \cite{PalffyPRA2007, GunstPRL2014, GunstPOP2015, WuPRL2018, GunstPRE2018}. We use a Multi-Configurational-Dirac-Fock method implemented in GRASP92 \cite{ParpiaCPC1996} for the relativistic bound electronic wavefunctions and numerical solutions to the Dirac equation with $Z_{\rm{eff}} = q$ for the free electrons under the single-active electron approximation. The nuclear reduced transition probability $B=3.5$ W.u. (Weisskopf units) for the 4.85 keV transition $IS \rightarrow T$ in $^{93}$Mo was taken from the model calculation in Ref.~ \cite{HasegawaPLB2011}.
We have checked the accuracy of our electronic matrix elements by reproducing existing experimental \cite{ensdf} or theoretical  internal conversion coefficients \cite{Roesel1978}. The agreement is on the level of $10$\%.

\begin{figure}
\centering
\includegraphics[width=1.0\linewidth]{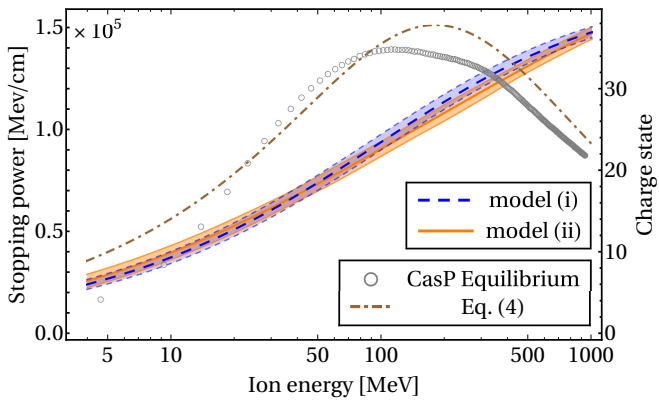}%
\caption{Right axis: Mean charge state  $\bar{q}$ (lines) and width $w$ (shaded area) for $^{93}$Mo ions as a function of ion energy using the fit formulas {\bf (i)} from Ref.~\cite{NikolaevPLA1968} (blue dashed line and blue shading) and {\bf (ii)} in Ref.~\cite{SchiwietzNIMPRSB2001} (orange solid line and orange shading). Left axis: Stopping power for $^{93}$Mo ions in the carbon solid target as a function of the total ion energy. Grey circles: CasP simulation considering an averaged equilibrium charge state. Brown dash-dotted line: the Javanainen semi-empirical formula \cite{JavanainenNIMPRSB2012}. A density of $2.0$ g$/$cm$^3$ for the carbon target was considered.
\label{fig:csd}}
\end{figure}

For  the ion charge distribution in the beam $f_{q}(E^{ \mathrm {ion}}_{q, \alpha})$ and the stopping power $-\left.\left( dE^{\mathrm {ion}}/dx \right)\right|_{E^{ \mathrm {ion}}_{q, \alpha}}$ we employ state-of-the-art models, empirical fits and software packages developed mostly by Schiwietz and Grande. For the charge state distribution we adopt for each ion energy a Gaussian distribution with mean charge state $\bar{q}$ and width $w$ defined as $w = \left[ \sum_{q'} (q' - \bar{q})^2 f_{q'} \right]^{1/2}$ \cite{BetzRMP1972}. The values $\bar{q}$ and $w$ can be obtained from multi-parameter least-square fits applied to a large collection of experimental data points. For our purposes we compare three different models: {\bf (i)} a general fitting formula introduced in 1968 by Nikolaev and Dmitriev \cite{NikolaevPLA1968}, {\bf (ii)} a multi-parameter  least-square fit by Schiwietz and Grande that has been applied to published solid-state data for 840 experimental data points \cite{SchiwietzNIMPRSB2001}, and  {\bf (iii)} an improved charge-state formula for $\bar{q}$ with asymptotic dependencies that include resonance effects and  shell-structure effects  in an iterative fitting procedure \cite{SchiwietzNIMPRSB2004}. For Mo channeling through a C foil, the calculated mean charge state using models {\bf (ii)} and  {\bf (iii)} are nearly identical, such that we use only model {\bf (ii)} in the following.  Figure~\ref{fig:csd} illustrates the good agreement of the mean charge state $\bar{q}$ and the width $w$ obtained for Mo ions using models {\bf (i)} and  {\bf (ii)} \cite{SchiwietzNIMPRSB2001,NikolaevPLA1968}.


We now proceed to calculate the stopping power of Mo ions and to evaluate the NEEC probabilities $P_{q}^{\alpha}$. Since Ref.~\cite{ChiaraNature2018} assumes that NEEC occurs in the C layer of the target, we consider the scenario that a
$^{93m}$Mo ion beam of energy 820 MeV traverses a C target of density $2.0$ g$/$cm$^3$ \cite{stopMaterial} and thickness approx. $100$ $\mu$m, sufficient to bring the ions to a full stop.  Upon deceleration, electrons recombine into the Mo ions in the available atomic vacancies depending on the ionic charge state. The considered energy interval allows NEEC  into the $L$, $M$ $N$ and $O$ shells and is larger than the one available in the experiment \cite{ChiaraNature2018}, thus providing an upper limit for the NEEC probability within the C target.

The stopping power was obtained using the state-of-the-art unitary-convolution-approximation stopping-power model implemented by Schiwietz and Grande in the Convolution approximation for swift Particles (CasP) code  \cite{Caspwebpage, SchiwietzPRA2011, GrandePRA1998, SchiwietzNIMPRSB1999, AzevedoNIMPRSB2000, GrandeNIMPRSB2002, GrandeNIMPRSB2009, SchiwietzNIMPRSB2012}. CasP takes ionization and electron capture processes into account and can provide both $-\left.\left( dE^{\mathrm{ion}}/dx \right)\right|_{E^{ \mathrm {ion}}_{q, \alpha}}$ for an ion of given charge $q$ at the resonance energy of interest in each possible channel $\alpha$ as well as an equilibrium charge-state-distribution averaged stopping power at a specific ion energy. For the total NEEC probability (\ref{total_P}) it is appropriate to consider the stopping power for specific charge states in the CasP calculation, henceforth denoted as CasP-q. However, we have also used the equilibrium CasP calculation to check the reliability of the stopping power results by comparison with the semi-empirical formula introduced by Javanainen \cite{JavanainenNIMPRSB2012},
\begin{equation}
  -\frac{d E^{\mathrm{ion}}}{dx} = \frac{4 \pi Z_p^2 e^4}{m_e v^2} N Z_t L,
\end{equation}
where $e$ is the electron charge, $v$ is the projectile velocity, $N$ is the atomic density of the target, $Z_p$ and $Z_t$ are the atomic numbers of the projectile and the target, respectively, and $L$ is the stopping number $L = 1.1209 \sqrt{1+0.2021 \ln{\left( Z_p/Z_t \right)}} \ln{\left( 1 + \kappa \chi \right)}$ with $k\approx 1.1229$ and $\chi = 0.2853 [Z_t^{1/2}/(1.2 Z_p^{1/3})] \xi$. Furthermore, $\xi=[(m_e v^3)/(Z_p \hbar \omega_0 v_0)]$ with $\hbar \omega_0$ being the mean excitation energy of the target electrons given by $\hbar \omega_0 = Z_t I_0 = 10 Z_t$ eV, and $v_0$ the Bohr velocity. The calculated stopping power in the carbon target as a function of the ion energy are compared for the case of Mo ions in Fig.~\ref{fig:csd}. For the CasP equilibrium calculation we have used 200 points with $50$ keV$/$u ion energy spacing from $50$ keV$/$u to $1000$ keV$/$u. Figure \ref{fig:csd} shows that both models deliver similar stopping powers. The CasP results underestimate the stopping power at low ion energies, a feature that  has been already addressed in Ref.~\cite{SigmundBook2014}.

\begin{figure}
\centering
\includegraphics[width=1.0\linewidth]{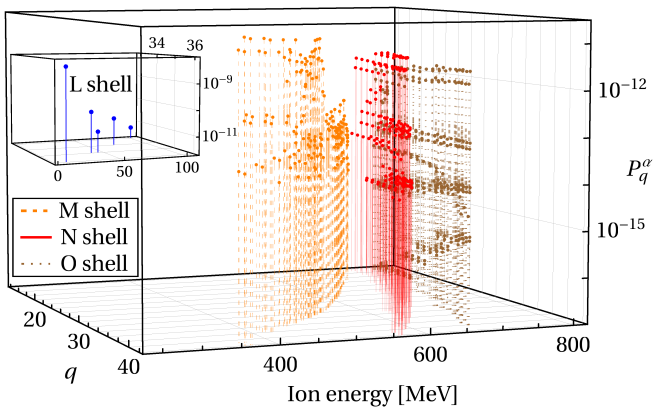}
\caption{NEEC probability $P_{q}^{\alpha}$ for all considered 648 recombination channels as a function of charge state $q$ and ion energy. For illustration  the $L$-shell values are presented in the inset.
\label{fig:pqalpha}}
\end{figure}

We obtain the partial  $P_{q}^{\alpha}$  NEEC probabilities into each possible channel combining the stopping power calculated with CasP-q and the corresponding NEEC resonance strength values according to Eq.~(\ref{total_P}). The results are shown in Fig.~\ref{fig:pqalpha}. We consider 648 NEEC channels for charge states from Mo$^{14+}$ to Mo$^{42+}$ and recombination into all possible orbitals of the $L$, $M$, $N$ and $O$ atomic shells. Among these, $23$ consider  NEEC into the respective electronic ground state and $625$ into excited states. This builds upon the smaller set of 333 NEEC cross sections calculated and presented in Refs.~\cite{WuPRL2018, GunstPRE2018} to consider all relevant channels for the present scenario. We note that NEEC into the $K$-shell is not possible due to the corresponding binding energy larger than $4.85$ keV. The lowest charge state  Mo$^{14+}$ occurs according to Fig.~\ref{fig:csd} only for very small ion energies, for which NEEC into the  free $N$ and $O$ shells is energetically forbidden. NEEC into these atomic shells can only proceed for higher ion energies and higher charge states, larger than 20+. The largest calculated $P_{q}^{\alpha}$ value is the one for NEEC into the  $L$-shell orbitals,  with  $P_{q}^{\alpha}\sim 10^{-9}$ for the $2p_{3/2}$ orbital. NEEC with recombination into the higher considered shells is less probable, with $P_{q}^{\alpha}$ values of approx. $10^{-11}$, $10^{-12}$, $10^{-12}$ for the $M$, $N$ and $O$ shells, respectively. Recombination in even higher shells is not possible  since the resonance condition would require that incoming ions have higher  energies  than considered here.

Summing over all the possible NEEC channels and including  the charge state distribution according to the second line of Eq.~(\ref{total_P}), we obtain the total probability $P$. The numerical results are shown in Table~\ref{table:ptotal} for the discussed stopping power models and the charge state distributions {\bf (i)} and  {\bf (ii)}. Using the CasP-q model and the charge state distribution {\bf (ii)}, we obtain $P=2.58\times10^{-11}$. This value changes only slightly when using the older charge state distribution model {\bf (i)}  ($P=2.66\times10^{-11}$).  We also present the NEEC probability calculated with the equilibrium charge state distributions from CasP calculations or the semi-empirical formula in Ref.~\cite{JavanainenNIMPRSB2012}. Among all calculated probabilities, the difference between the largest and the smallest value is only 17\%, confirming that all considered combinations of models predict a much smaller NEEC probability than the observed isomer depletion probability.


\renewcommand{\arraystretch}{1.25}
\begin{table}
  \centering
  \footnotesize
  \begin{tabular}{lcc}
  \hline\hline
  &  & \tabularnewline[-0.4cm]
  \diagbox[width=6em]{$-dE/dx$  }{ $q$  } & {\bf (i)} \cite{NikolaevPLA1968} &
  {\bf (ii)} \cite{SchiwietzNIMPRSB2001} \tabularnewline
  
  &  &  \tabularnewline[-0.4cm] \hline
  &  &  \tabularnewline[-0.4cm]
  CasP-q & $2.66 \times 10^{-11}$ & $2.58 \times 10^{-11}$  \tabularnewline
  CasP Equilibrium & $2.73 \times 10^{-11}$ &$2.54 \times 10^{-11}$ \tabularnewline
  Ref.~\cite{JavanainenNIMPRSB2012} & $2.43 \times 10^{-11}$  & $2.26 \times 10^{-11}$  \tabularnewline
  &  & \tabularnewline[-0.4cm] \hline\hline
  \end{tabular}
  \caption{The total NEEC   excitation probability $P$ for  $^{93m}$Mo via the $4.85$ keV $E2$ transition for the discussed stopping power  and charge state distribution models. }
  \label{table:ptotal}
\end{table}
\renewcommand{\arraystretch}{1}


Table~\ref{table:pshell} presents the individual contribution from each atomic shell to the probability of NEEC excitation.  Surprisingly, although  $P_{q}^{\alpha}$ is largest for the $L$ shell, the corresponding total NEEC probability is negligible.
The reason is that the fraction of ions $f_{q}(E^{ \mathrm {ion}}_{q, \alpha})$ with high charge states and $L$-shell vacancies is vanishingly small for the ion energy required by the resonant condition.  We  note that the  $L$-shell contributions  in Table~\ref{table:pshell}  have  a one order of magnitude difference between the two charge state distribution models. This is expected because the charge state required for the capture into the $L$ shell is far away from the averaged charge state at the resonance condition \cite{BetzRMP1972, DmitrievJETP1965}. However, this discrepancy does not affect the total NEEC probability, which is determined by recombination into the $M$, $N$ and $O$ shells. The largest contribution to the NEEC probability is from the capture into the $M$-shell channels, for which all predictions are  very close to each other on the few percent level, regardless of the chosen charge state distribution model.


\renewcommand{\arraystretch}{1.25}
\begin{table}
  \centering
  \footnotesize
  \begin{tabular}{lcccc}
  \hline\hline
  & & & & \tabularnewline[-0.4cm]
 $q$ model & $L$ shell & $M$ shell & $N$ shell & $O$ shell \tabularnewline
  & & & &  \tabularnewline[-0.4cm] \hline
  & & & &  \tabularnewline[-0.4cm]
{\bf (i)} & $5.68 \times 10^{-18}$ & $1.53 \times 10^{-11}$ & $7.54 \times 10^{-12}$ & $3.77 \times 10^{-12}$ \tabularnewline
  
 {\bf (ii)} & $1.48 \times 10^{-19}$ & $1.47 \times 10^{-11}$ & $7.34 \times 10^{-12}$ & $3.73 \times 10^{-12}$ \tabularnewline

  & & & & \tabularnewline[-0.4cm] \hline\hline
  \end{tabular}
  \caption{The NEEC excitation probability for each individual capture shell.
  The stopping power was calculated with the charge-selective  CasP-q code. }
  \label{table:pshell}
\end{table}
\renewcommand{\arraystretch}{1}


The nine orders of magnitude discrepancy between the theoretical NEEC probability calculated in this work and the experimental excitation probability in Ref.~\cite{ChiaraNature2018} sheds doubts on whether NEEC was the process behind the observed isomer depletion. We note  that our predictions should be considered as upper limits for the excitation probability, since
  in the experiment  the ions are expected to have only energies between approx. 600 MeV and 300 MeV at the presumed NEEC site. This energy interval covers the largest contribution from recombination into the $M$ shell, but only partially the ones of other shells. Since in the experimental target  the C layer was backed by a stopping Pb layer, we have calculated also the NEEC probability for a 820 MeV ion beam channeling and coming to a full stop through a Pb target of density 11.35 g/cm$^{3}$ \cite{stopMaterial}. We find  $P\approx 5\times 10^{-11}$ using the stopping power model \cite{JavanainenNIMPRSB2012} and charge state distribution {\bf (ii)}. This value is likely an overestimate since for Pb our approximations for the recombining electrons are more inaccurate. Therefore, the NEEC probability in the experimental target with C and Pb layers considering smaller ion energies than we have assumed for our calculation should remain on the order of $P\approx  10^{-11}$ or less. This order of magnitude corroborates with the results obtained for a laser-plasma-based NEEC scenario for  $^{93m}$Mo isomer depletion where recombination into the $L$-shell had a seizable contribution and $P\approx 10^{-10}$ \cite{WuPRL2018}.

Which process could be responsible for the observed excitation? Ref.~\cite{ChiaraNature2018} presents estimates of Coulomb excitation and inelastic scattering probabilities which yield approx.~$ 10^{-6}$ for the Pb and C targets, respectively. These values, although too small to explain the observed excitation, are much larger than the calculated NEEC probability. Since our theoretical results have shown that for NEEC only a very small ion energy interval in the deceleration process  contributes and the strongest channels are suppressed, it is not surprising that  non-resonant nuclear excitation processes  may be more efficient.  It remains an open question whether the observed excitation can be related to a novel channel  so far disregarded in state-of-the-art theory.



\bibliographystyle{apsrev-no-url-issn.bst}
\bibliography{neecbrefs}{}

\end{document}